\documentclass[aps,prl,twocolumn,showpacs,superscriptaddress,amsmath,amssymb]{revtex4-1}

\usepackage{xcolor}
\usepackage{graphicx}
\usepackage{subfigure}
\usepackage{multirow}
\usepackage{booktabs}
\usepackage{colortbl}
\usepackage{hhline}
\usepackage{siunitx}
\usepackage[left,modulo]{lineno}
\usepackage{upgreek}
\usepackage{braket}
\usepackage[
			colorlinks=true,
			urlcolor=blue,
			linkcolor=blue,
			citecolor=blue,
			filecolor=blue
			pagebackref=blue,
			]{hyperref}

\begin{document}

\title{Visible Out-of-plane Polarized Luminescence and Electronic Resonance from Black Phosphorus}

\author{L\'eonard Schu\'e}
\affiliation{D\'epartement de Chimie, Universit\'e de Montr\'eal, Montr\'eal, Qu\'ebec, Canada}
\affiliation{D\'epartement de Physique, Universit\'e de Montr\'eal, Montr\'eal, Qu\'ebec, Canada}
\author{F\'elix A. Goudreault}
\affiliation{D\'epartement de Physique, Universit\'e de Montr\'eal, Montr\'eal, Qu\'ebec, Canada}
\author{Ariete Righi}
\affiliation{Departamento de F\`{i}sica, Universidade Federal de Minas Gerais, Belo Horizonte, Brasil}
\author{Geovani C. Resende}
\affiliation{Departamento de F\`{i}sica, Universidade Federal de Minas Gerais, Belo Horizonte, Brasil}
\author{Val\'erie Lefebvre}
\affiliation{D\'epartement de Chimie, Universit\'e de Montr\'eal, Montr\'eal, Qu\'ebec, Canada}
\author{\'Emile Godbout}
\affiliation{D\'epartement de Chimie, Universit\'e de Montr\'eal, Montr\'eal, Qu\'ebec, Canada}
\author{Marcos A. Pimenta}
\affiliation{Departamento de F\`{i}sica, Universidade Federal de Minas Gerais, Belo Horizonte, Brasil}
\author{Michel C\^ot\'e}
\affiliation{D\'epartement de Physique, Universit\'e de Montr\'eal, Montr\'eal, Qu\'ebec, Canada}
\author{S\'ebastien Franc\oe{}ur}
\email{sebastien.francoeur@polymtl.ca}
\affiliation{D\'epartement de G\'enie Physique, \'Ecole Polytechnique de Montr\'eal, Montr\'eal, Qu\'ebec, Canada}
\author{Richard Martel}
\email{r.martel@umontreal.ca}
\affiliation{D\'epartement de Chimie, Universit\'e de Montr\'eal, Montr\'eal, Qu\'ebec, Canada}

\begin{abstract}

Black Phosphorus (BP) is unique among layered materials owing to its homonuclear lattice and strong structural anisotropy. While recent investigations on few layers BP have extensively explored the in-plane ($a,c$) anisotropy, much less attention has been given to the out-of-plane direction ($b$). Here, the optical response from bulk BP is probed using polarization-resolved photoluminescence (PL), photoluminescence excitation (PLE) and resonant Raman scattering along the zigzag, out-of-plane, and armchair directions. PL reveals an unexpected $b$-polarized emission occurring in the visible at 1.75 eV, far above the fundamental gap (0.3 eV). PLE indicates that this emission is generated through $b$-polarized excitation at 2.3 eV. The same electronic resonance is observed in resonant Raman scans, where the scattering efficiency of both $A_{g}$ phonon modes is enhanced. These experimental results are fully consistent with DFT calculations of the permittivity tensor elements and demonstrate the remarkable extent to which the anisotropy influences the optical properties and carrier dynamics in black phosphorus.

\end{abstract}

\maketitle

\section{\label{sec:Intro}Introduction}

\begin{figure*}[hbtp]
 \includegraphics[scale=1.5]{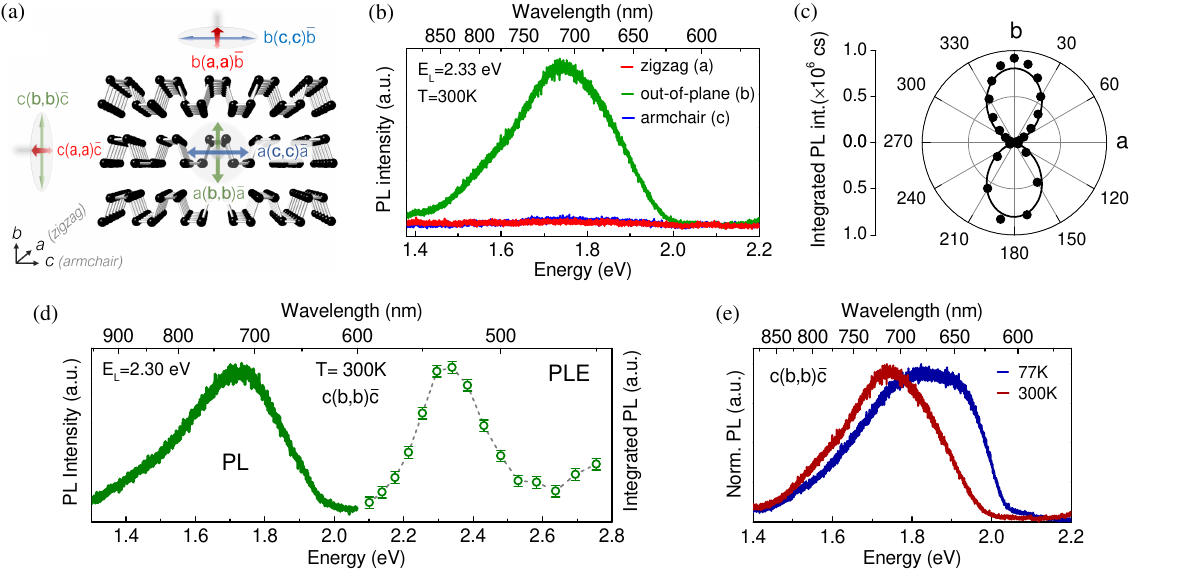}
  \caption{\label{fig_1}\textbf{Highly polarized visible luminescence in bulk BP.} (a) Schematic illustration of the polarization-resolved spectroscopy experiments. All polarization configurations are identified using the Porto notation. (b) PL spectra recorded with the parallel configuration along the zigzag ($a$, red), out-of-plane ($b$, green) and armchair ($c$, blue) axes under 2.33 eV excitation. (c) Integrated PL intensity as a function of the polarization direction. The solid curve represents the emission profile from a linear dipole oriented along $b$. (d) Left: Photoluminescence spectrum recorded in the $c(b,b)\bar{c}$ configuration using 2.30 eV excitation. Right: Excitation energy dependent profile of the PL intensity integrated over the 1.4-2.0 eV emission range. (e) Normalized PL spectra recorded under $b$-polarized excitation at \SI{300}{\kelvin} (dark red) and \SI{77}{\kelvin} (dark blue).}
\end{figure*}

Black Phosphorus (BP) is a low-symmetry semiconductor with remarkable electrical \cite{Xia2014,Qiao2014,Li2014,Koenig2014}, optical \cite{Xia2014,Low2014,Tran2014,Qiao2014,Buscema2014}, mechanical, \cite{Jiang2014,Fei2014-2,Hu2014,Wei2014} and thermal \cite{Fei2014,Jain2015,Jang2015} properties. As illustrated in Fig. \ref{fig_1}(a), BP presents an orthorhombic crystal structure consisting of layers stacked together in the out-of-plane direction ($b$). The in-plane lattice is puckered and shows two non-equivalent directions, defined by alternated atomic positions in the so-called zigzag ($a$) and armchair ($c$) directions. Because of the weak interaction between layers, mono as well as few layers BP have been prepared since the first exfoliation report in 2014 \cite{Li2014} and thoroughly investigated experimentally and theoretically. Most importantly, these studies have shown that the layer thickness can be used to tune the material's band gap \cite{Low2014,Tran2014,Zhang2014,Cai2014} with optical transitions that can be pushed from the mid-infrared (MIR) in the bulk crystal \cite{Li2016,Chen2019,Carre2021} up to the visible in a monolayer BP \cite{Li2016,Yang2015,Surrente2016}. 

Since BP exhibits anisotropy in both in-plane and out-of-plane directions, a thorough understanding requires resolving properties along the three nonequivalent crystal directions. The strong in-plane anisotropy was quickly noted in early studies on bulk crystals \cite{Asahina1984} and more recently in few-layers BP samples \cite{Qiao2014,Ribeiro2015,Kim2015-2,Li2016,Mao2016,Phaneuf2016}. Most notably, the luminescence at the fundamental band gap is strongly polarized along the armchair direction for both bulk and 2D crystals \cite{Tran2014,Wang2015,Li2016,Chen2019}. 
External modulation using applied uniaxial strains \cite{Wang2015-2,Li2017} or electric fields \cite{Xia2014,Kim2015} have further highlighted this anisotropic response of few layers BP. The presence of highly polarized Raman resonances in bulk BP at high energies in the visible spectrum \cite{Ling2016,Wang2018,Mao2019} demonstrated that the anisotropy is manifest over the whole electronic band structure. Hence, optical and electrical anisotropies are not limited to the band gap edges. The state of knowledge on in-plane anisotropy contrasts with that involving the out-of-plane direction \cite{Sugai1981,Ikezawa1983,Sugai1985,Qiao2014,Zhu2020,Lin2020}. An important result is the relatively lower out-of-plane electron-hole effective masses with respect to in-plane directions \cite{Takao1981,Asahina1982,Narita1983}. More recently, a remarkably high electrical conductivity was measured in a BP/graphene vertical field-effect transistor \cite{Kang2016}. These are some of the very few measurements where the out-of-plane anisotropy is probed \cite{Akahama1983,Jang2015,Wang2016,Sun2017}.

In this work, we use polarization-resolved photoluminescence and resonant Raman scattering along the zigzag, out-of-plane and armchair directions to thoroughly investigate the band structure anisotropy of bulk BP in the visible range of the optical spectrum. Our measurements reveal, for the first time, a room-temperature luminescence occurring at 1.75 eV, which is strongly polarized in the out-of-plane direction. This above band gap emission is surprising as it violates Kasha's rule, which favors light emission from the lowest energy states \cite{Kasha1950}. The luminescence is shown to be intrinsic to BP crystals and not from its oxides. To elucidate the origin of this unexpected luminescence, we have systematically investigated the related band structure electronic states using photoluminescence excitation (PLE) and resonant Raman scattering over an excitation range from 1.9 to 2.7 eV. The results reveal an unreported resonance at $\sim$ 2.3 eV in both the PL and Raman response. Polarization selection rules and density functional theory (DFT) calculations of the complex dielectric permittivity are used to elucidate the origin of the strongly polarized resonance.


\section{\label{sec:Results}Results and Discussion}

\subsection{\label{sec:Luminescence}Luminescence}

Figure \ref{fig_1}(b) presents the luminescence measured from bulk BP at \SI{300}{\kelvin}. The sample has a thickness of \SI{10}{\micro\meter} and it is excited at \SI{2.33}{\electronvolt} with a power of \SI{500}{\micro\watt}. Using a parallel excitation-detection polarization, the luminescence is acquired along all three crystallographic directions ($a$, $b$, $c$) and polarizations indicated in Fig. \ref{fig_1}(a). A complete description of the experimental setup can be found in section S1 of the Supplementary Information. Despite having a fundamental gap of about 0.3 eV, all BP samples investigated revealed a luminescence in the visible range, which is discussed next. 


Consistent with the extensive literature on BP, no significant luminescence is detected in Fig. \ref{fig_1}(b) across the visible and near-infrared (NIR) regions (1.2 eV $<E<$ 3 eV) 
 along the zigzag ($b(a,a)\bar{b}$, red) and armchair directions ($b(c,c)\bar{b}$, blue). 
 In contrast, a luminescence signal is readily measured from the out-of-plane ($b$) polarization. The $c(b,b)\bar{c}$ spectrum (green) in Fig. \ref{fig_1}(b) presents a broad and asymmetric emission located at \SI{1.75}{\electronvolt}. All measurements confirm that the emission is independent of the measurement axis: no significant signal can be measured in the $c(a,a)\bar{c}$ and $a(c,c)\bar{a}$ configurations, and the $b$-polarized luminescence can be detected in the $a(b,b)\bar{a}$ configuration as well (see Fig. \ref{fig_PLall}). Since this emission is isotropic in the $ac$ polarization plane, the emission dipole appears to be strictly oriented along the $b$ axis. This is indeed confirmed by the emission intensity profile as a function of polarization in the $ab$ plane shown in Fig. \ref{fig_1}(c). The polarization contrast ($I_b/I_a$) of $\sim 50$ is approaching that measured from the fundamental bandgap (at \SI{0.3}{\electronvolt}) of $I_c/I_a\sim 100$ \cite{Chen2019}. Interestingly, this luminescence can only be observed in a parallel configuration, with excitation and emission polarizations aligned with the $b$ axis. Similar measurements were repeated on several samples originating from large crystals from different commercial sources (see Fig. \ref{fig_SE-HQ}). Hence, this visible emission at \SI{1.75}{\electronvolt} is ubiquitous to the bulk crystal and, for the reasons discussed below, it is ascribed to an intrinsic electronic process in BP.

\begin{figure*}[hbtp]
\includegraphics[scale=1.25]{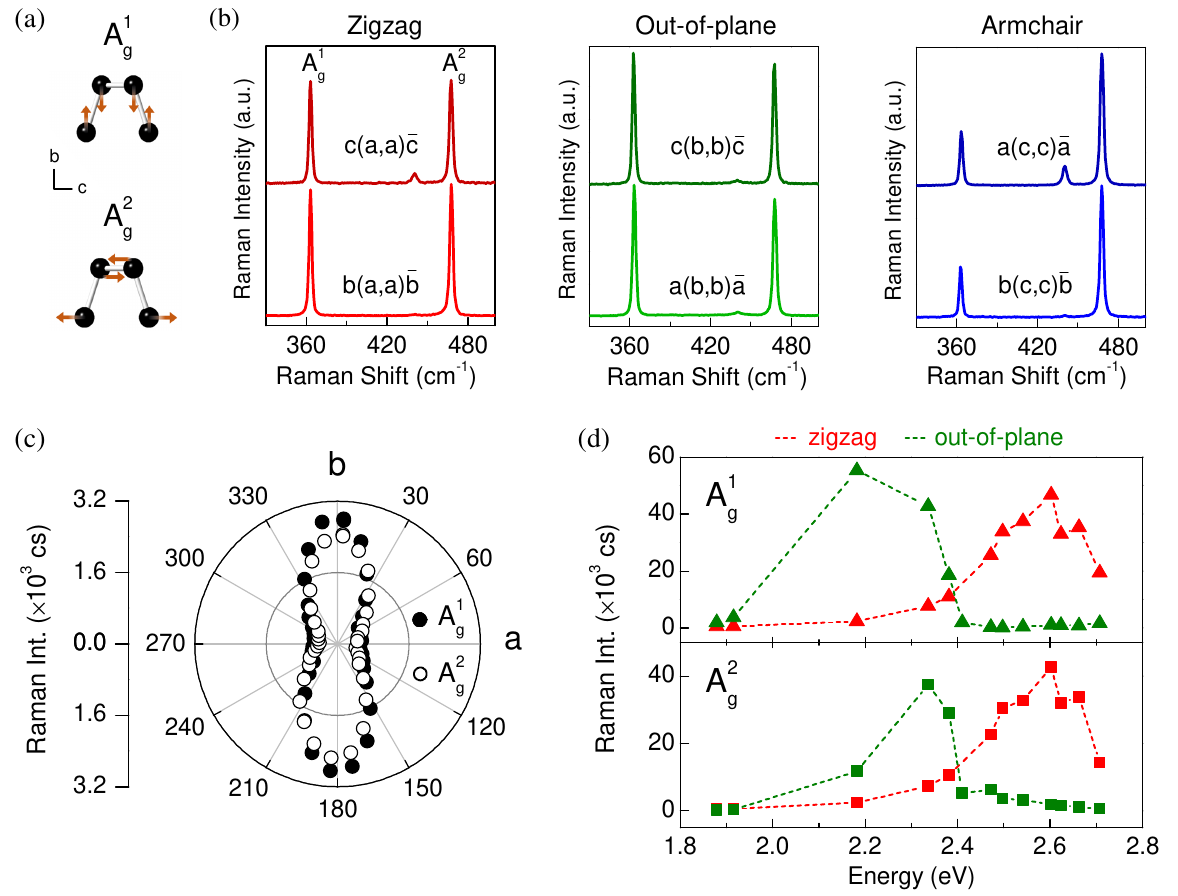}
  \caption{\label{fig_Raman}\textbf{Anisotropy of the BP band structure revealed by Raman scattering.} (a) Atomic displacements of the $A_{g}^{1}$ (top) and $A_{g}^{2}$ (bottom) phonon modes in BP. (b) Raman spectra under \SI{2.33}{\electronvolt} excitation recorded along $a$, $b$ and $c$ under two distinct configurations. (c) Integrated Raman intensity of the $A_{g}^{1}$ (full circles) and $A_{g}^{2}$ (empty circles) as a function of the polarization angle in the $ab$ plane. Black solid lines show the fitting curves. (d) Energy-dependence profiles of the normalized Raman intensity of $A_{g}^{1}$ (top) and $A_{g}^{2}$ (bottom) recorded along the zigzag (red) and out-of-plane (green) directions in the $ab$ polarization plane.}
\end{figure*}

Figure \ref{fig_1}(d) shows that the emission intensity strongly depends on the excitation energy and polarization. On the right, the integrated PL intensity is plotted as a function of the excitation energy (PLE). As expected from the continuum of excited states over the whole excitation range studied (\SIrange{2.1}{2.8}{\electronvolt}), some residual luminescence from hot carriers is always observed. However, a strong enhancement occurs at \SI{2.33}{\electronvolt}, which corresponds to the most common laser wavelength used in optical measurements. The absorption is strongly polarized along $b$, as no luminescence can be detected for $a$- or $c$-polarized excitation. These absorption and emission bands are separated by a Stokes shift of about \SI{0.58}{\electronvolt}.

The low-temperature PL shown in Fig. \ref{fig_1}(e) indicates that the maximum PL energy blueshifts by $\sim$\SI{100}{\milli\electronvolt} when lowering the sample temperature down to \SI{77}{\kelvin}. Interestingly, this trend is opposite to the redshift reported for the $c$-polarized band gap emission \cite{Chen2019} but similar to the usual shift observed from most semiconductors. The luminescence intensity increases quasi-linearly without shifting over more than two orders of magnitude of excitation powers (from 20 $\mu$W to 2500 $\mu$W), suggesting that the emission occurs through a bulk-like density of states (see Fig. \ref{fig_PvsI}). The emission is stable and does not change over extended periods of time. For example, the emission at \SI{1.75}{\electronvolt} is almost unchanged after 6 months in air, and does not increase or decrease when subject to surface degradation and photo-oxidation \cite{Favron2015} (see Fig. \ref{fig_Stab}). In contrast to recent studies \cite{Lu2015-PL,Gan2015,Zhao2018,Nan2021}, the observed emission does not match the characteristics of the emission from oxidized by-products. Indeed, the BP oxides luminescence is rather detected at \SI{1.85}{\electronvolt} (Fig. \ref{fig_Ox}) and it is unpolarized (Fig. \ref{fig_Ox}(a)). Furthermore, this oxide emission is emitted along all three axes ($a$, $b$, and $c$), which contrasts with the highly polarized luminescence described here. Finally, it does not shift with temperature (Fig. \ref{fig_Ox}(c)). From all of the evidences, we suggest that the observed visible luminescence and the absorption band at 2.33 eV are intrinsic to BP. As is demonstrated next, the resonant Raman scattering of $b$-polarized transitions in this spectral region further supports this interpretation.

\begin{figure*}[ht]
 \includegraphics[scale=1.35]{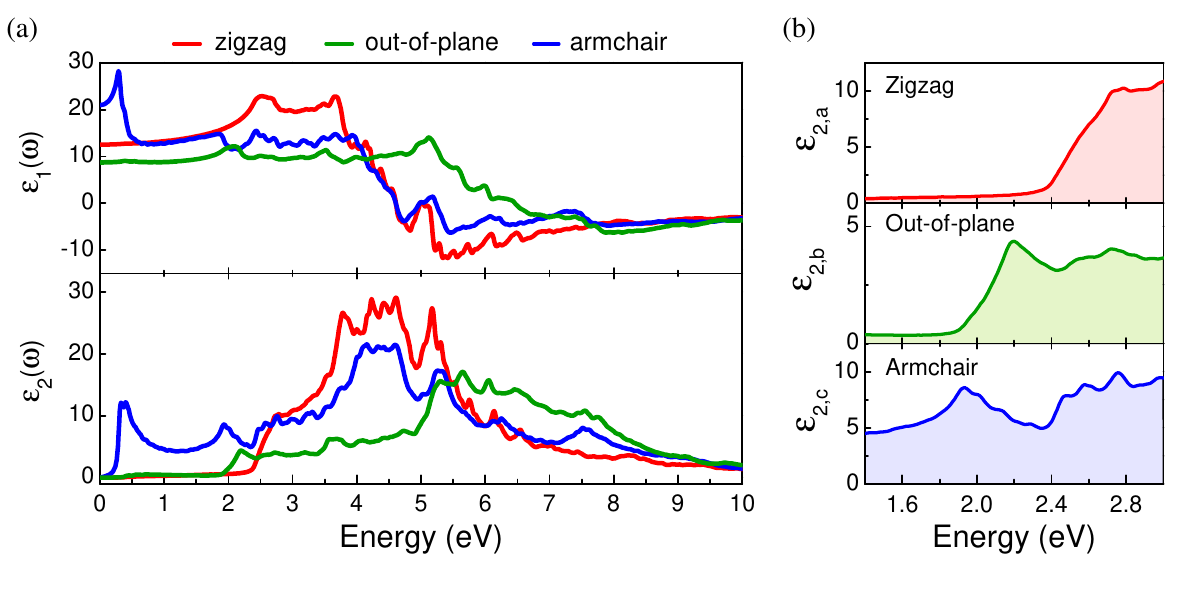}
  \caption{\label{fig_Eps} \textbf{Theoretical calculations of the dielectric response of bulk BP.} (a) DFT-calculated real and imaginary parts of the complex dielectric function of bulk BP along the zigzag (red), out-of-plane (green) and armchair (blue) directions. (b) Zoom in of the imaginary part of the dielectric function in the visible region.}
\end{figure*}

\subsection{\label{sec:Raman}Raman scattering}

Figure \ref{fig_Raman}(b) presents representative Raman spectra of the two $A_{g}$ phonon modes (illustrated in panel (a)) measured using a parallel configuration with the polarization aligned along $a$, $b$ and $c$ directions. For each polarization, the spectra correspond to the two associated optical axes. As can be seen, the Raman response obtained using an excitation at \SI{2.33}{\electronvolt} is mostly determined by the polarization of the light and it is not sensitive to the measurement axes (see also Fig. \ref{fig_SpRam}). In contrast to the $a$-polarized signal, where both A$_{g}$ phonons present similar intensity, the intensity of the A$_{g}^{2}$ mode dominates in the polarized $c$ direction. This anisotropy is typical of BP and often used to identify the crystallographic orientation of BP crystals, provided that one takes into account the natural birefringence and dichroism of BP and interference effects in the sample structure \cite{Kim2015,Ling2016,Phaneuf2016}.

Under $b$-polarized excitation, the spectra appear quite similar to that in the $a$ direction. Nonetheless, the integrated intensity as a function of polarization angle in the $ab$ plane, as shown in Figure \ref{fig_Raman}(c), reveals that the signal is strongly polarized along the $b$-axis for both phonon modes. This enhancement is ascribed below to an electronic resonance.

Figure \ref{fig_Raman}(d) plots the Raman intensity as a function of the laser excitation energy, obtained using 13 different laser excitation energies in the range from 1.9 to \SI{2.7}{\electronvolt}, for both $a$ (red) and $b$ (green) polarizations (all spectra are presented in Fig. \ref{fig_SpRam_2}). Under $a$-polarized excitation, the Raman intensity of both modes starts to increase at about \SI{2.3}{\electronvolt} and reaches a maximum at \SI{2.6}{\electronvolt}, as previously reported \cite{Wang2018,Mao2019}. This is in stark contrast with the strong exaltation observed at 2.2-2.3 eV for both modes under $b$ polarization. This Raman resonance profile is, to our knowledge, reported here for the first time and indicates the presence of a strong $b$-polarized resonance in bulk BP. In previous studies, peculiar Raman scattering intensities of the A$_{g}$ modes under 532 nm excitation were found in bilayer BP \cite{Favron2015} and with defect modes in few-layers BP flakes \cite{Favron2018}, which appear connected somehow to the bulk resonance reported here. Because its energy position, linewidth and polarization selection rule agree with the PLE results presented above (Fig. \ref{fig_1}(d)), an electronic resonance in the band structure of BP is most likely at the origin of all of these phenomena.

\subsection{\label{sec:DFT}DFT calculations}

To gain insight into the nature of the electronic states of bulk BP and associated electron-photon interactions, \textit{ab initio} calculations of the complex dielectric function were performed along each of the three main directions. Figure \ref{fig_Eps}(a) displays the energy-dependent real ($\epsilon_{1}$, top) and imaginary ($\epsilon_{2}$, bottom) parts for polarizations along $a$ (red), $b$ (green) and $c$ (blue) directions. In agreement with the early calculations of Asahina \textit{et al.} \cite{Asahina1984}, the permittivity strongly depends on polarization up to about \SI{9}{\electronvolt}.

Starting from the fundamental bandgap of bulk BP (0.3 eV), dipolar transitions are only allowed under $c$-polarized excitations up to about 2 eV. As can be seen from the imaginary part of the permittivity shown in Fig. \ref{fig_Eps}(b), the thresholds for $b$- and $a$-polarized transitions are limited to the visible range. While the $c$-polarized transitions involve a broad continuum of conduction and valence band states, the extinction coefficient for $a$-polarized transitions rapidly increases from 2.4 eV, as previously reported \cite{Tran2014,Qiao2014}. A similar behavior is observed for the $b$ polarization, where a significant threshold can be identified at 1.9 eV. This implies a rapidly growing joint density of states that can participate in electronic transitions. Most interestingly, an extremum is easily identified at 2.3 eV. This energetic structure at 2.3 eV precisely corresponds to the optical resonance observed in both in PL excitation (Fig. \ref{fig_1}(d)) and Raman scattering (Fig. \ref{fig_Raman}(d)).


In intrinsic semiconductors, electron-photon interaction generally occurs near the energy where the joint optical density of states becomes important. For the fundamental gap at 0.3 eV, the bands involved are readily assigned to valence and conduction bands extrema. However, such identification is much less trivial for higher energy resonances in $a$ and $b$ polarizations. In particular, pocket-like regions and band nesting phenomena have been shown to significantly influence the optical response and carrier dynamics \cite{Carvalho2013}. Furthermore, such high-energy resonances may likely involve contributions from several distinct regions of the electronic band-structure. For these reasons, the identification of the electronic bands involved as well as the location of the excitation/recombination pathways will require further theoretical investigation.

Thermalization of excited carriers in molecules and solids generally occurs on time-scales much shorter than that for radiative emission. This is the basis of the so-called Kasha's rule \cite{Kasha1950}, which states that emission predominantly occurs at the lowest available excited states. Hence, a visible luminescence occurring at 1.75 eV is rather unexpected from a material such as BP whose fundamental band gap is only 0.33 eV (300K) \cite{Chen2019}. Although all evidences shown here suggest intrinsic emission from bulk BP, the origin of this unusual PL remains to be further confirmed. For instance, the experimental configuration used here (with the measurement axis parallel to the basal plane) cannot be easily adapted to completely rule out possible contributions from edge electronic states \cite{Liang2014}. However, this luminescence is strongly polarized along the out-of-plane direction irrespective of the edge studied ($a$ or $c$), and can only be generated with out-of-plane excitation, which makes rather clear that the crystal anisotropy plays a key role. Therefore, it is possible to speculate that a fraction of the carriers generated by a $b$-polarized excitation conserves their symmetry representation on a time-scale comparable of, or longer than, the radiative emission lifetime. That is, the relaxation of carriers creates a bottleneck at the lowest available states compatible with this symmetry representation because of a weak coupling with the lowest $c$-polarized fundamental gap states. The pronounced directional response, as illustrated in Fig. \ref{fig_Eps}, suggests that BP behaves like three different materials when probed according to each possible optical axis.

To date, Visible and NIR luminescence emission in BP is generally assigned to thin crystals of few layers. Although the bulk PL emission energy reported here appears similar with that of the monolayer \cite{Yang2015,Xu2016,Pei2016,Li2017,Tan2018,Wang2021}, the dipole orientation is clearly distinct : the fundamental gap of BP always remains $c$-polarized irrespective of the sample thickness. Moreover, the DFT-calculated dielectric response of monolayer BP for out-of-plane polarization shows that dipole transitions occur at much higher energy, above 3 eV (see Fig. \ref{fig_bulk-mono}). In any case, further insights on the interband transitions behind this visible resonance in bulk BP will undoubtedly provide precious information on the few-layers optical properties. Besides, the PL emission is often used to analyze exfoliated samples in solutions \cite{Hanlon2015,Kang2016-2} and to confirm their few-layers thickness. A visible luminescence is also reported in BP quantum dots \cite{Niu2016}, where all of the three high symmetry directions are randomly oriented along the measurement axis and thus can contribute to the optical response. The bulk emission reported here therefore calls for more care in the characterization of BP crystals in solution, in particular when using photoluminescence spectroscopy.

\section{\label{sec:Ccl}Conclusion}

In conclusion, we observed a room-temperature visible luminescence in bulk BP, which is highly polarized along the out-of-plane direction of the crystal. PLE experiments have uncovered the presence of a strong $b$-polarized emission resonance around 2.3 eV responsible for the PL emission detected at 1.75 eV. Highly sensitive to polarization, excitation energy, and temperature, this luminescence is intrinsically related to the anisotropy of the electronic band structure of bulk BP. Polarization and energy-resolved Raman spectroscopy further reveals a significant enhancement of the Raman scattering efficiency of the A$_{g}$ phonon modes due to excited states in the same energy range and with the same polarization dependence. \textit{Ab initio} calculations of the dielectric permittivity tensor of bulk BP finally show an onset of dipole-allowed transitions along the out-of-plane direction in the visible range above 2 eV. These findings shed light on the huge anisotropy in black phosphorus and offer new insights into its fundamental properties.

\section{\label{sec:Met}Methods}

\paragraph{Sample preparation.} Bulk BP flakes (thickness $\geq$ 1 $\mu m$) were mechanically exfoliated using PDMS stamps from bulk crystals of two different sources (Smart Elements, HQ Graphene) and transferred onto Si/SiO$_{2}$ substrates.

\paragraph{Photoluminescence and Raman spectroscopy.} BP samples were excited in a backscattering geometry. The polarization of incident and emitted/scattered lights was controlled independently using a rotating half-wave plate and fixed polarizers. Near-diffraction limited resolution was achieved using a 50X objective minimizing the electric field polarized along the optical axis. All spectra were measured in the parallel configuration (labeled in Figure \ref{fig_1} using the Porto notation) where the incident laser polarization is parallel to the emitted/scattered light polarization. The samples were mounted on a two-axis rotation stage (\textit{see Supplementary Information}) such that the three crystallographic directions, zigzag ($a$), out-of-plane ($b$) and armchair ($c$), can be oriented along the measurement axis. Photoluminescence excitation measurements were performed using a supercontinuum laser source with a tunable laser line filter.
Energy-dependent Raman experiments were performed using a T64000 Jobin-Yvon Horiba triple-monochromator spectrometer equipped with a N$_{2}$-cooled CCD detector and a 1800 g/mm grating that provides a resolution better than 1 cm$^{-1}$. The BP samples were excited with 13 laser lines covering the visible range from 1.92 to 2.73 eV (Ar/Kr) and at 1.88 eV (solid state laser). The Si Raman peak at 521.6 cm$^{-1}$ was used to calibrate the Raman shift. The laser power was kept below 500 µW to avoid sample heating and photo-induced oxidation \cite{Favron2015}. The crystalline orientation of the samples was adjusted by minimizing the signal of the $B_{g}$ modes.

\paragraph{Raman intensity correction factor.}The measured Raman intensity in bulk BP is mostly altered by birefringence and linear dichroism due to the different complex refractive indices along the zigzag, armchair and out-of-plane directions. To determine the intrinsic Raman response, the intensity correction factor ($F$) was calculated for the three principal axes ($a$, $b$, $c$) following a method reported previously \cite{Yoon2009} and detailed in \textit{Supporting information}.

\paragraph{Dielectric tensor calculation.} Materials dielectric properties were calculated in the framework of density functional theory (DFT). Calculations were carried out using the \texttt{abinit} software \cite{Gonze2020} which implements DFT within a plane wave basis. Ground state calculations were done using a PBE exchange and correlation functional \cite{perdew_generalized_1996} with a plane wave energy cutoff of 25Ha. A $20\times 20\times 20$ $\boldsymbol{k}$-point mesh generating more than 2400 points in the irreducible Brillouin Zone (BZ) was employed. We used a scalar relativistic pseudopotential provided by the PseudoDojo project \cite{van_setten_pseudodojo:_2018}. The frequency-dependent dielectric tensor was computed within perturbation theory as developed by Sharma and Ambrosch-Draxl \cite{Sharma2004} whose formalism is implemented in the \texttt{optic} utility of \texttt{abinit} \cite{Gonze2016}. The tensor was calculated using a fine $60\times 60\times 60$ $\boldsymbol{k}$-point mesh in the full BZ computed in a non-self-consistent manner from the ground state calculation described above. Since band gaps obtained with DFT and the PBE functional are known to underestimate the experimental band gaps, to better compare to experiments, the band gap calculated was fixed to the experimental value \cite{li_direct_2017}.

\section{Acknowledgments}

Computations were executed on the B\'{e}luga cluster provided by Calcul Québec (www.calculquebec.ca) and Compute Canada (www.computecanada.ca). A.R., G.R. and M.P. acknowledge the financial support from the Brazilian agencies CNPq, FINEP, CAPES, FAPEMIG, and Brazilian Institute of Science and Technology in Carbon Nanomaterials (INCT)-Nanocarbono. This work was supported by the Natural Sciences and Engineering Research Council of Canada (NSERC) under grants RGPIN-2019-06545 (R.M.), RGPAS-2019-00050 (R.M.), RGPIN-2016-06666 (M.C.) and the NSERC GreEN Network, NETGP 508526-17 (R.M.). Support from the Canada Research Chairs (R.M.) and the Canada Foundation for Innovation (FCI, S.F. and R.M.) is acknowledged.

\newpage

\section{References}

\bibliography{Main}

\clearpage


\onecolumngrid

\section{\\
\vspace{9cm}
Supplementary Information: \\
\vspace{1cm}
Visible Out-of-plane Polarized Luminescence and Electronic Resonance from Black Phosphorus}

\renewcommand{\thefigure}{S\arabic{figure}}
\renewcommand{\thetable}{S\arabic{table}}
\setcounter{figure}{0}

\newpage

\section{Experimental set-up}

The $a$-polarized response (incident and scattered beam polarizations along $a$) is recorded under the $b(a,a)\bar{b}$ and $c(a,a)\bar{c}$ configurations, respectively, while the $b$-polarized and $c$-polarized spectra are acquired under the $a(b,b)\bar{a}$, $c(b,b)\bar{c}$ and $a(c,c)\bar{a}$, $b(c,c)\bar{b}$ configurations, respectively.

\begin{figure}[ht]
 \includegraphics[scale=0.6]{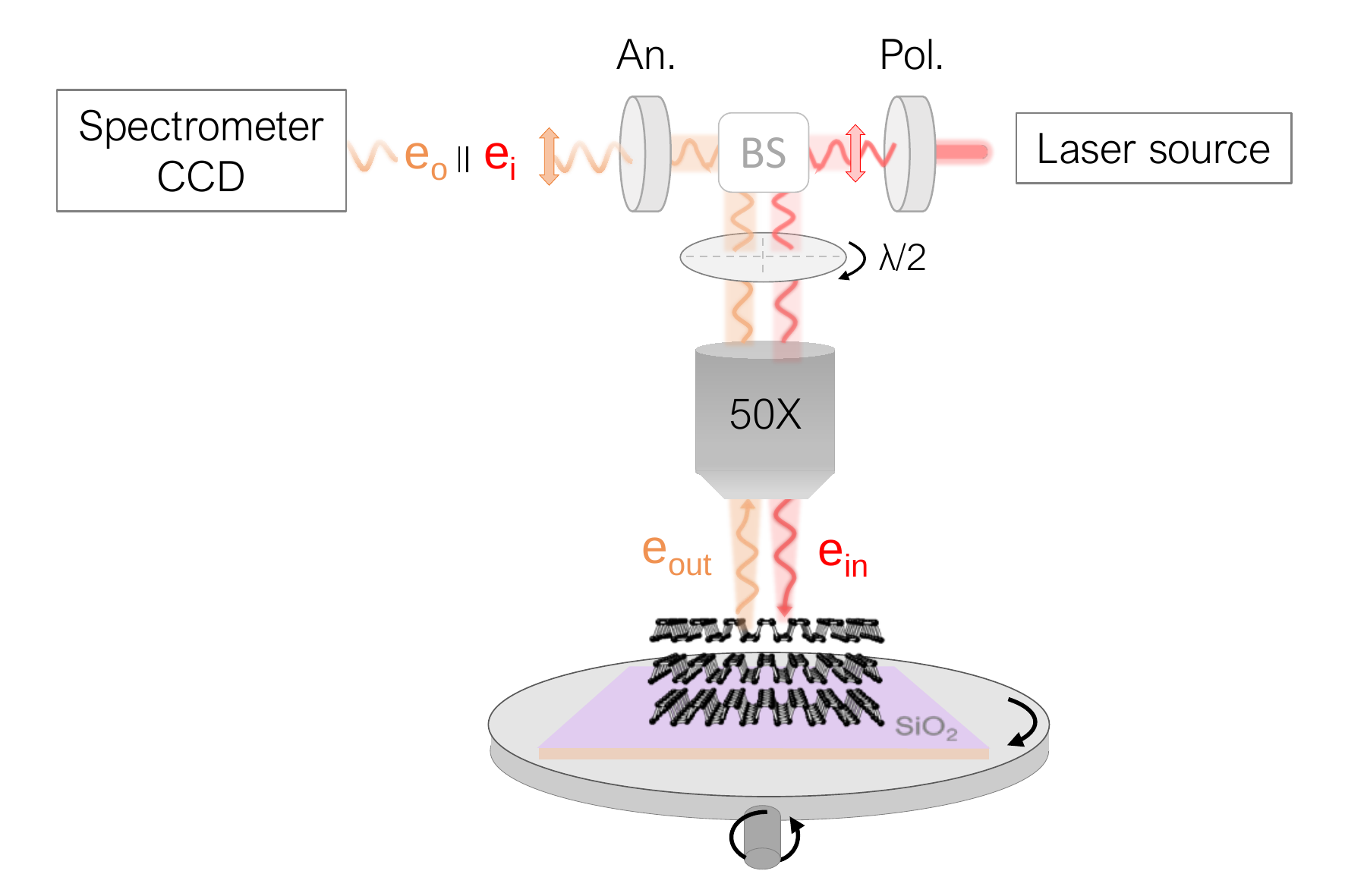}
  \caption{\label{fig_Mont}\textbf{Polarization-resolved Raman spectroscopy in bulk BP along the main crystallographic directions.} Schematic illustration of the polarization-resolved photoluminescence/Raman experiments, $e_{in}$ and $e_{out}$ are the incident and outgoing beam polarizations, respectively.}
\end{figure}

\section{PL spectra under all configurations}

\begin{figure}[ht]
 \includegraphics[scale=1.48]{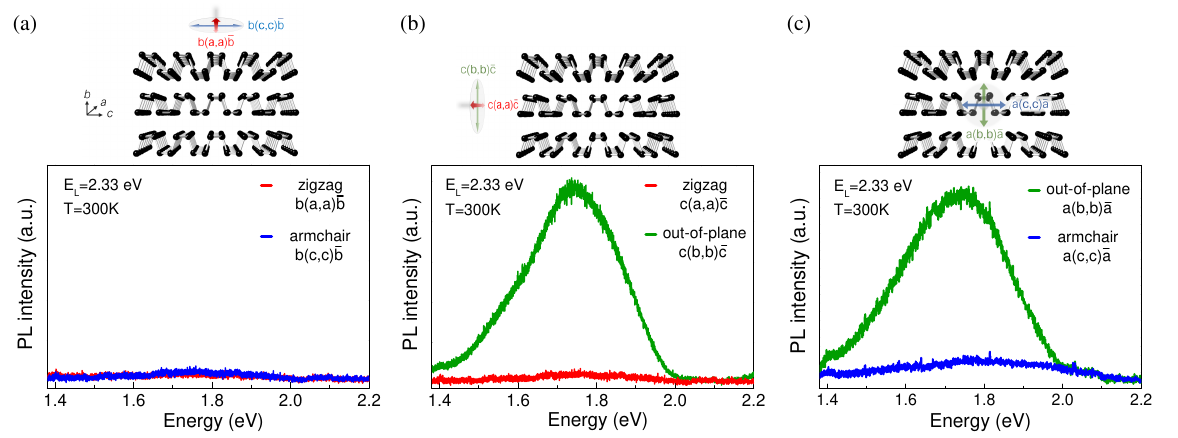}
  \caption{\label{fig_PLall}Photoluminescence spectra recorded under 2.33 eV excitation in the (a) $ac$, (b) $ab$ and (c) $ac$ polarization plane, respectively. T=300K.}
\end{figure}

\newpage

\section{Comparison of two different BP sources}

\begin{figure}[ht]
 \includegraphics[scale=1.28]{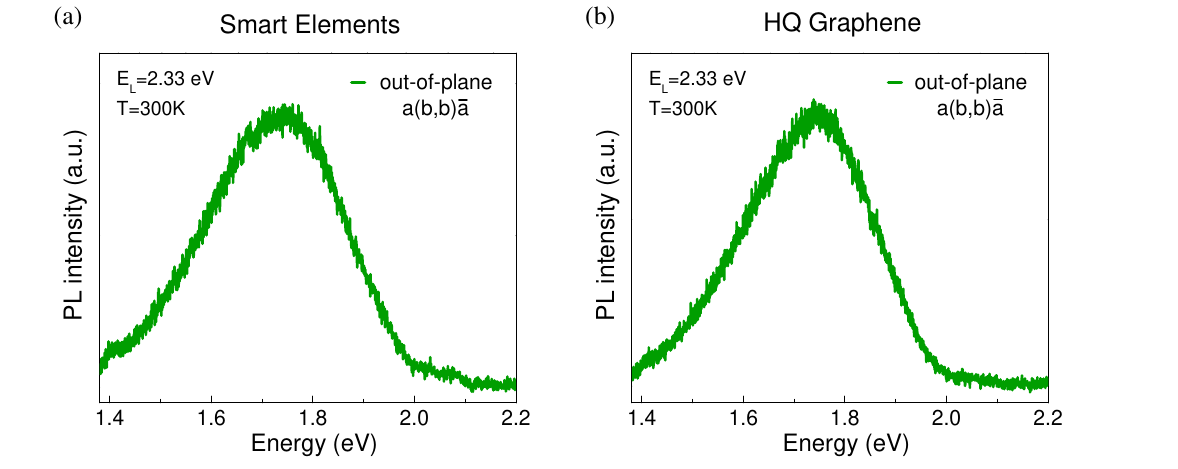}
  \caption{\textbf{Observation of a visible PL emission on two different BP sources.} Normalized PL spectra recorded on a bulk BP crystal exfoliated from the (a) Smart Elements and (b) HQ Graphene sources under out-of-plane polarized excitation. T=300 K; $E_{L}$=2.33 eV.}
  \label{fig_SE-HQ}
\end{figure}

\section{Low-temperature PL experiments}

\begin{figure}[ht]
 \includegraphics[scale=1.2]{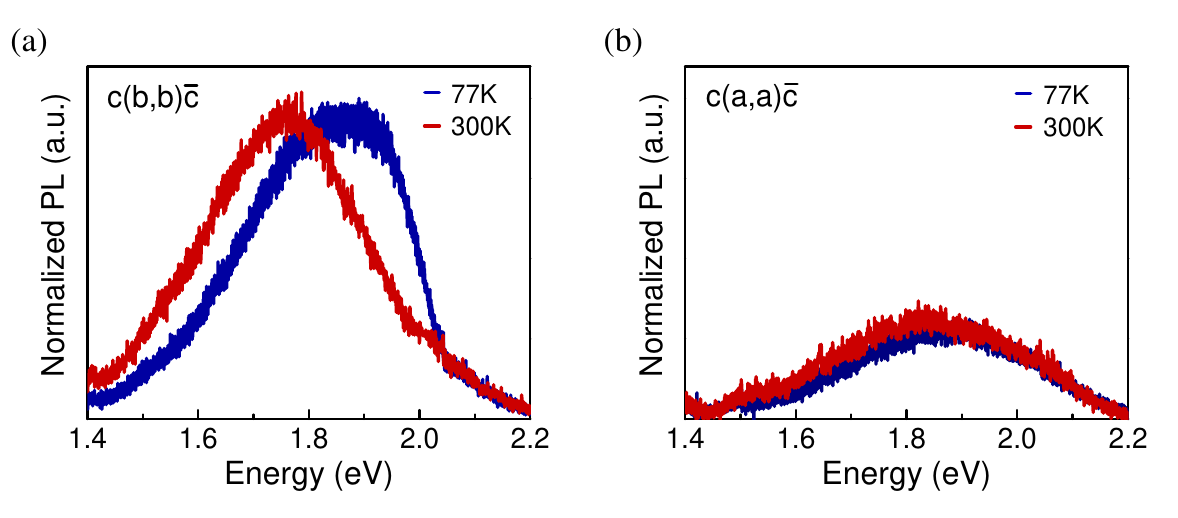}
  \caption{\label{fig_T-PL} Normalized PL spectra recorded under (a) out-of-plane and (b) zigzag polarized excitation at 300K (red) and 77K (blue).}
\end{figure}

At low-temperature (77K), the PL response shows very distinct behaviors within the $ab$ polarization plane. Along $b$, the PL band shifts at higher energy upon temperature decrease while it remains unchanged from 300K to 77K along the zigzag direction. This clearly indicates that the two signal are of different origins. The emission observed under $a$-polarized excitation mainly comes from oxide degradation species and shows no dependence on the polarization or the temperature as depicted in Figures \ref{fig_Stab}-\ref{fig_Ox}.

\newpage

\section{Power-dependence of the PL intensity}

\begin{figure}[ht]
 \includegraphics[scale=1.4]{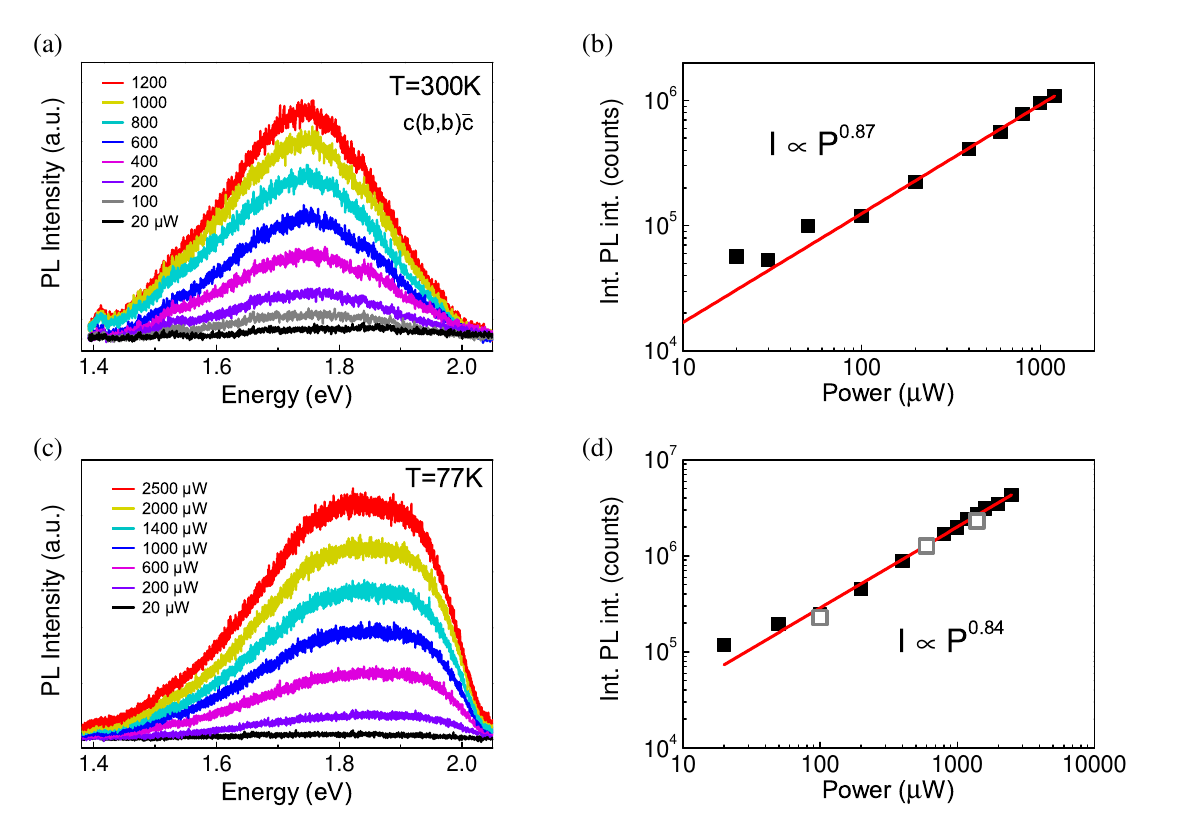}
  \caption{\textbf{Power-dependence of the PL emission of bulk BP.} (a) (c) PL spectra recorded along $b$ under various excitation powers at 300K and 77K. (b) (d) Plot (log-log scale) of the integrated PL intensity as a function of the excitation power at 300K and 77K. $E_{L}$=2.33 eV.}
  \label{fig_PvsI}
\end{figure}

\newpage

\section{PL emission stability in air}

\begin{figure}[ht]
 \includegraphics[scale=0.35]{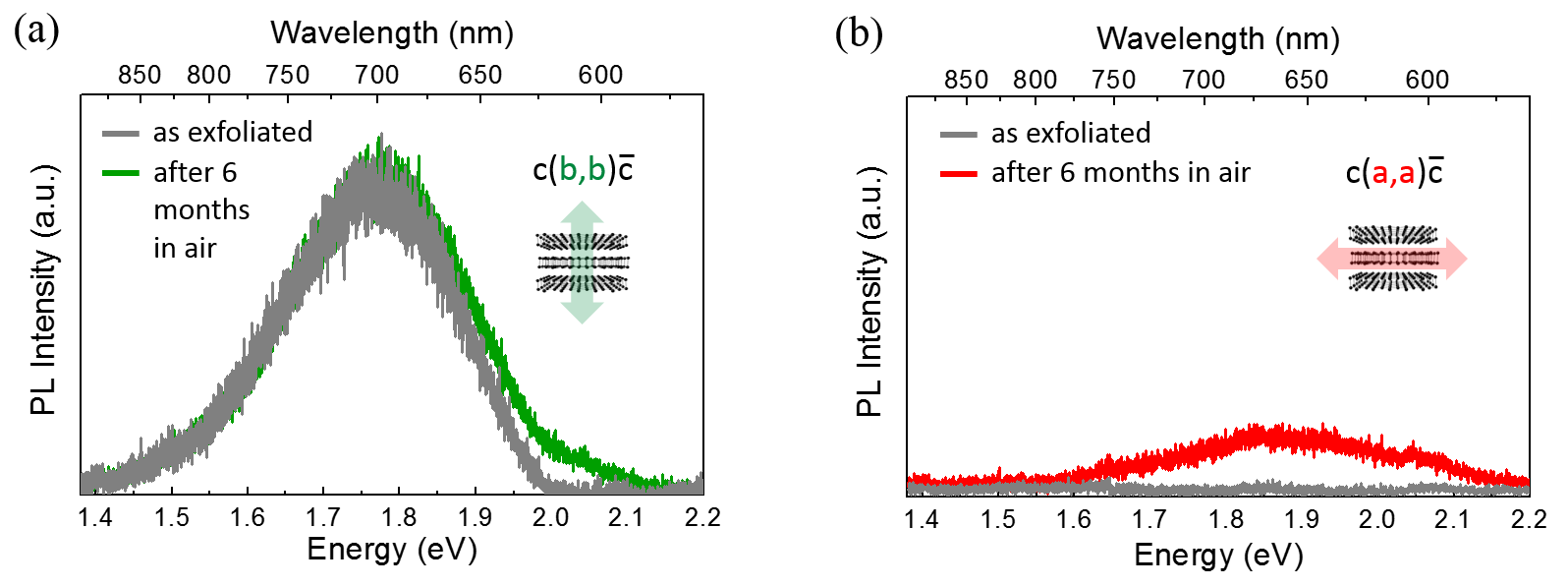}
  \caption{\textbf{Air stability of the PL emission of bulk BP.} PL emission spectra recorded on a bulk BP crystal right after the exfoliation (gray) and after 6 months in air (red) under (a) out-of-plane and (b) zigzag polarization. $E_{L}$=2.33 eV; T=300K.}
  \label{fig_Stab}
\end{figure}

\noindent
Experiments done on BP flakes left in air have revealed the strong stability of the PL emission which remains unchanged after 6 months (Fig.\ref{fig_Stab}(a)) while the oxide-related luminescence has increased (Fig.\ref{fig_Stab}(b)).

\section{PL characterization of highly oxidized BP surface}

\begin{figure}[ht]
 \includegraphics[scale=1.49]{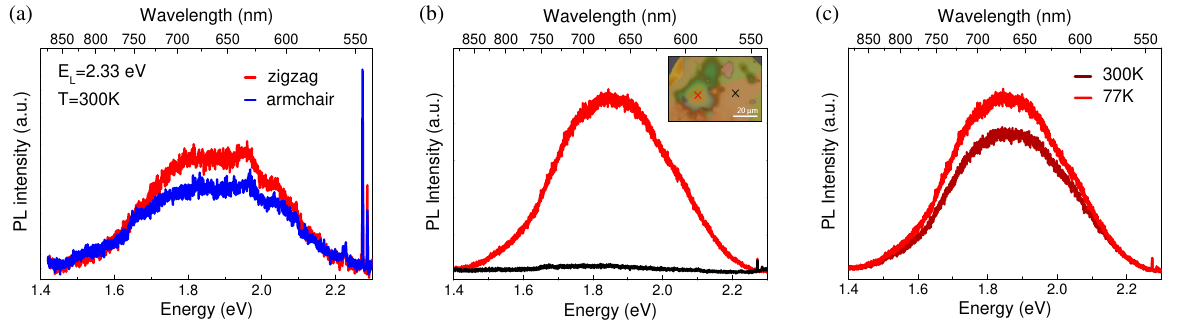}
  \caption{\textbf{Characterization of the oxide-related PL signal.} a) PL spectra recorded at 300K under zigzag (red) and armchair (blue) polarization excitation on a BP flake surface exposed to air. b) Inset: Optical image of another mechanically exfoliated BP flake after few hours exposition in air. PL spectra recorded at 77K on the highly degraded area of the flake (red) and next to it (black). c) PL spectra measured on the oxidized area at 77K (red) and 300K (dark red). $E_{L}$=2.33 eV.}
  \label{fig_Ox}
\end{figure}

\noindent
On highly degraded flakes such as the one shown in Figure \ref{fig_Ox}(b)(c), a strong PL emission is detected at higher energy around 1.85 eV (675 nm) which is no longer observed when exciting on a relatively clean part of the BP surface. This defect-related PL does not show any significant changes at low temperature in contrast to the out-of-plane polarized emission (Fig.\ref{fig_T-PL}).

\newpage

\section{Anisotropy of the Raman response in bulk BP}

\begin{figure*}[ht]
 \includegraphics[scale=1.5]{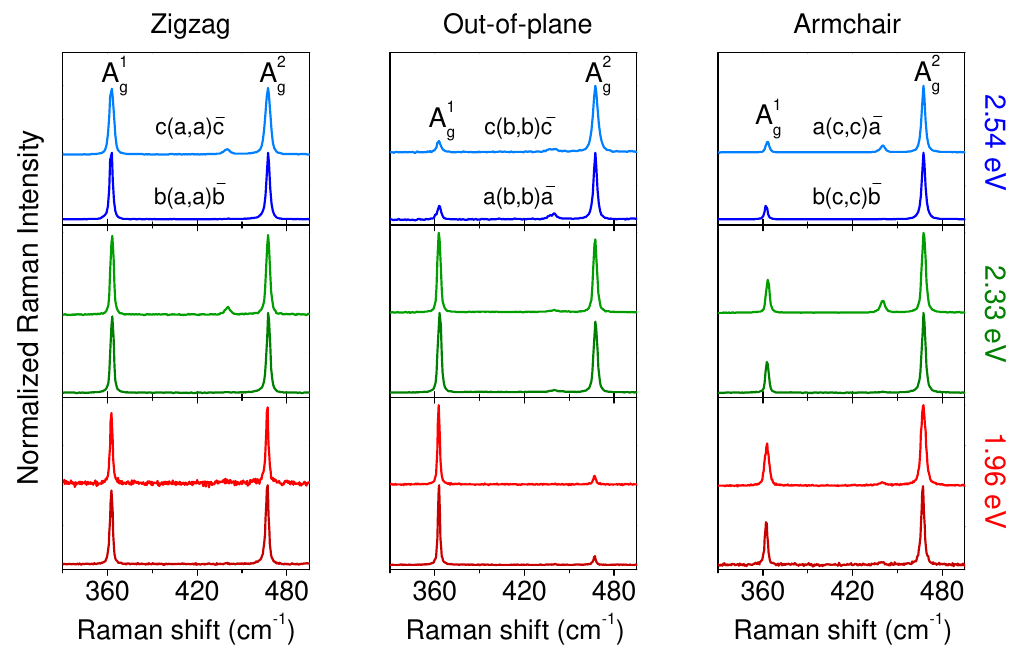}
  \caption{\label{fig_SpRam}\textbf{Polarization- and excitation energy dependence of the Raman of bulk BP.} Normalized Raman spectra of bulk BP recorded with excitation energies at 1.96 eV (632.8 nm), 2.33 eV (532 nm) and 2.54 eV (488 nm) in parallel configuration along the zigzag ($a$), out-of-plane ($b$) and armchair ($c$) directions.}
\end{figure*}

\noindent
All Raman spectra used for identifying the resonance are presented in Figure \ref{fig_SpRam} as a function of the polarization for three excitation energies. For each direction of the polarization, two different configurations were measured such as $c(a,a)\bar{c}$ (light curve) and $b(a,a)\bar{b}$ (dark curve) for the $a$-polarized response. In all cases, the A$_{g}$ peaks showed no dependence on the configuration used, indicating no contribution of edge phonons or other extrinsic effects from the edges of the sample \cite{Ribeiro2016}.

\newpage

\section{Anisotropy of the Raman response in bulk BP}

\vspace{1cm}

\begin{figure}[htbp]
 \centering
  \includegraphics[scale=0.48]{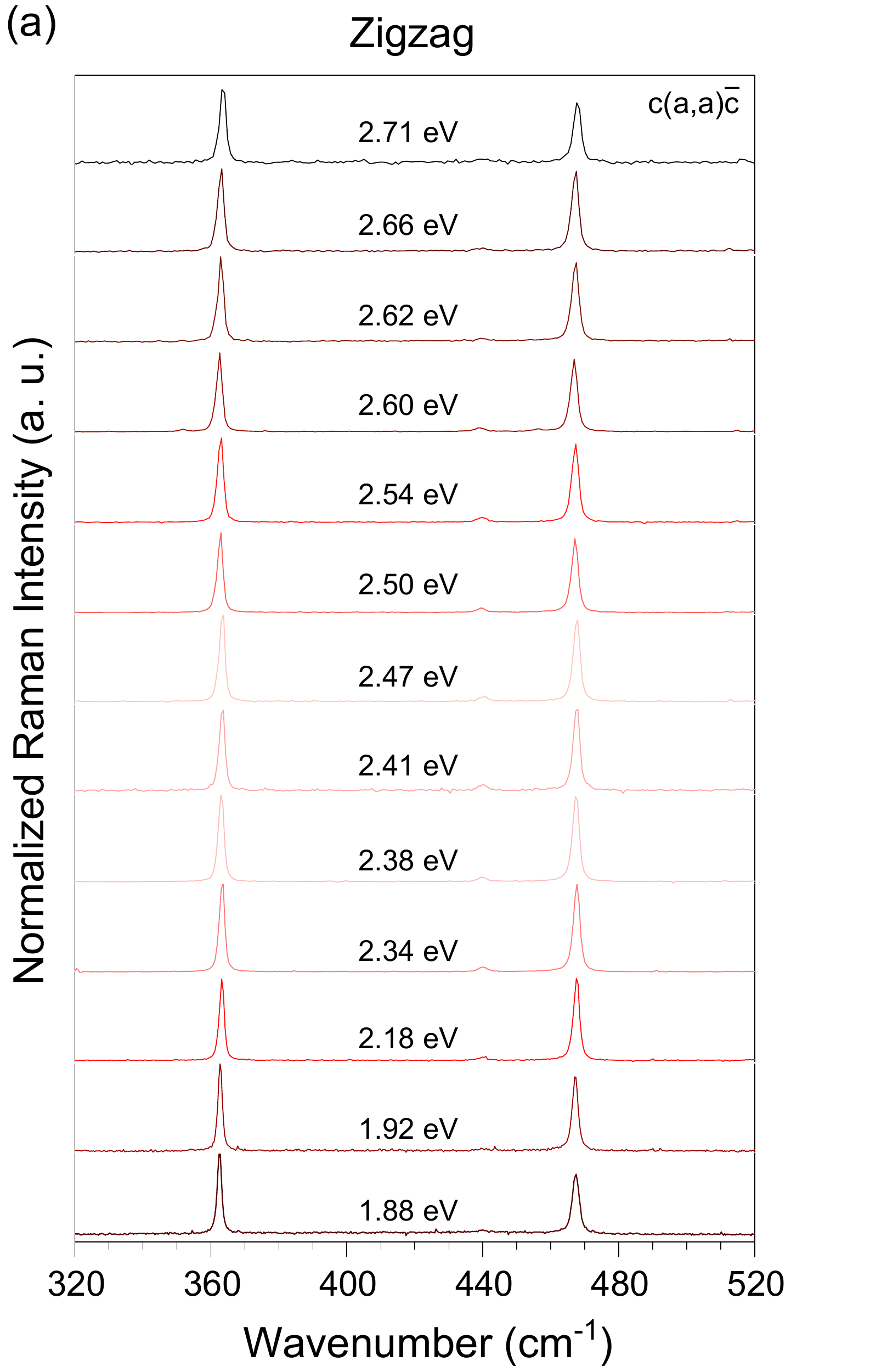}
 \centering
  \includegraphics[scale=0.483]{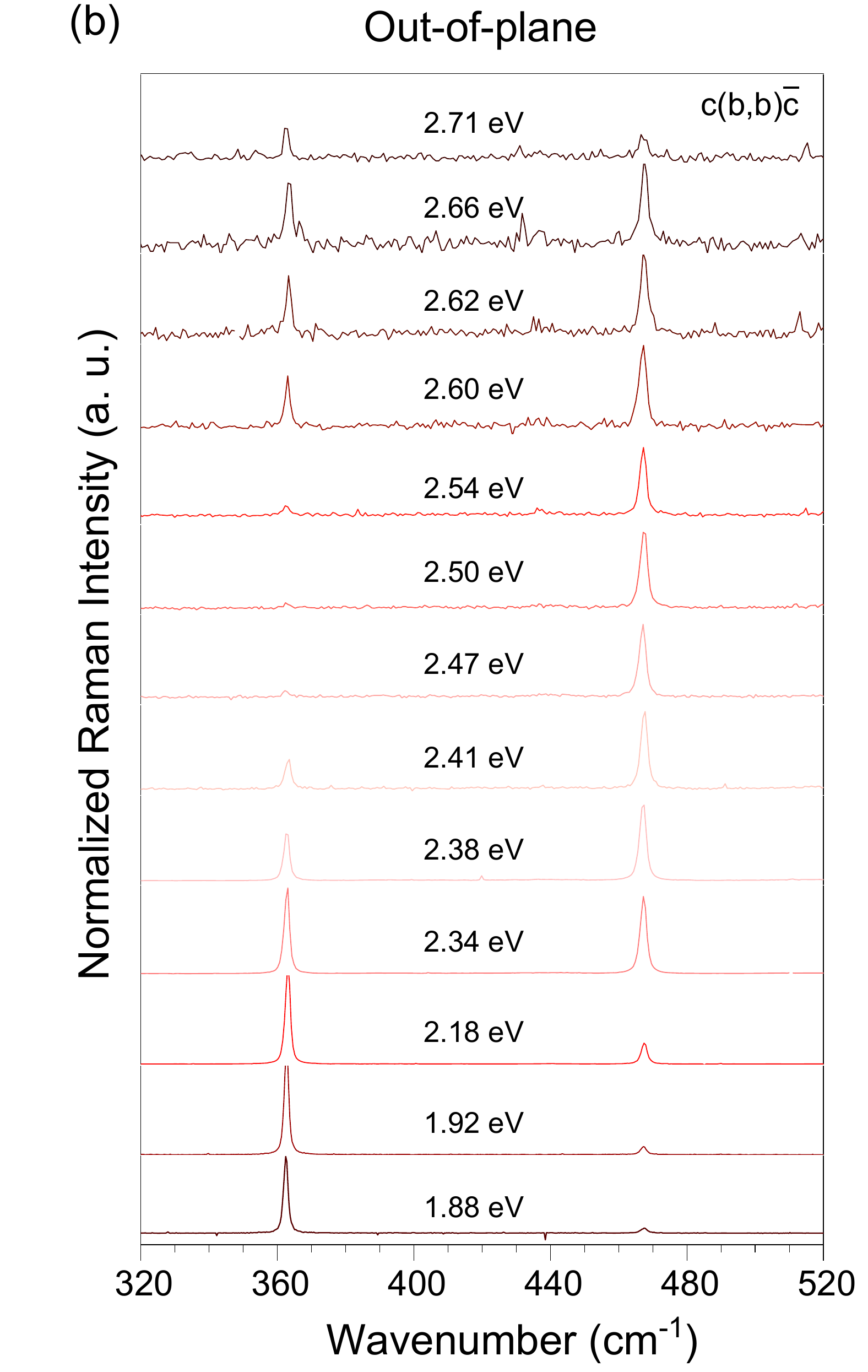}
   \caption{\textbf{Excitation-energy dependence of the Raman response of bulk BP.} Normalized Raman spectra of bulk BP recorded with excitation energies at 1.88 eV (660 nm), 1.92 eV (647.2 nm), 2.18 eV (568.2 nm), 2.34 eV (530.8 nm), 2.38 eV (520.8 nm), 2.41 (514.5 nm), 2.47 eV (501.7 nm), 2.50 eV (496.5 nm), 2.54 eV (488 nm), 2.60 eV (476.5 nm), 2.62 eV (472.7 nm), 2.66 eV (465.8 nm) and 2.71 eV (458 nm) in parallel configuration for (a) zigzag ($c(a,a)\bar{c}$) and (b) out-of-plane ($c(b,b)\bar{c}$) polarizations.}
  \label{fig_SpRam_2}
\end{figure} 

\newpage

\section{Comparison of Bulk and Monolayer permittivity tensors}

\begin{figure}[ht]
 \includegraphics[scale=0.8]{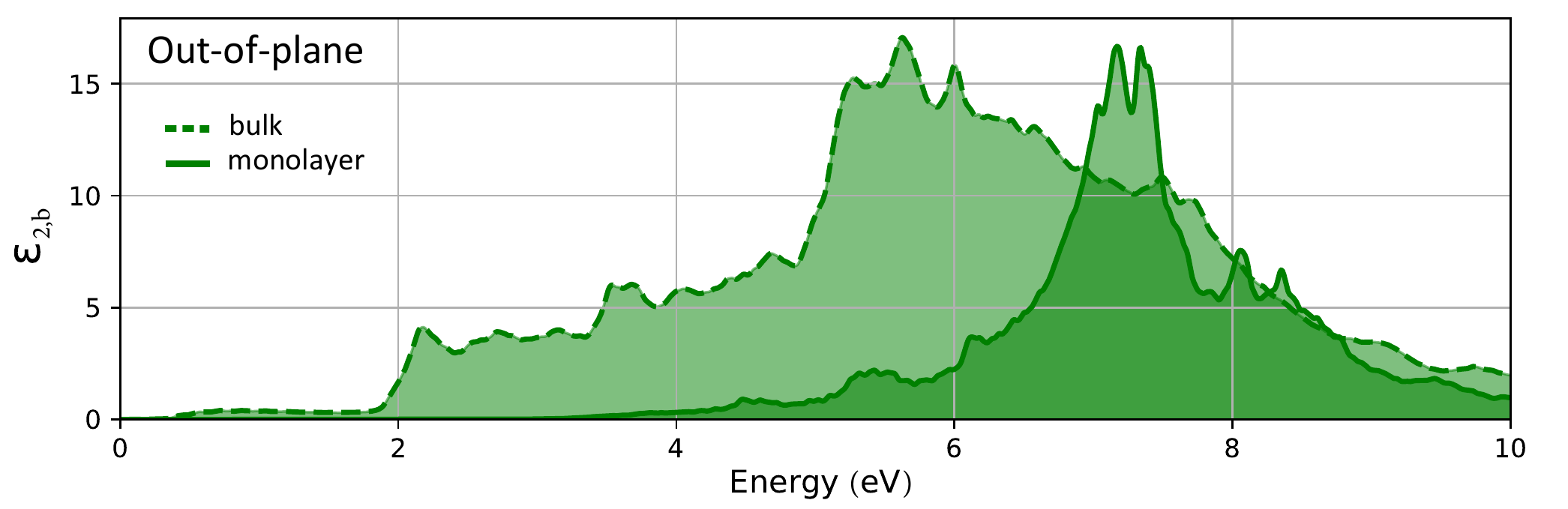}
  \caption{\label{fig_bulk-mono} DFT-calculated imaginary part of the complex dielectric function of bulk (dashed curve) and monolayer (full curve) BP along the out-of-plane direction.}
\end{figure}

\newpage

\section{Complex refractive indices in bulk BP}

\vspace{1cm}

To measure the influence of the extrinsic effects on the Raman intensity, the energy-dependent real (Figure \ref{fig_Eps}(a), top) and imaginary (Figure \ref{fig_Eps}(b), bottom) parts of the complex dielectric function $\tilde{\epsilon}$ were calculated by summation over all valence/conduction bands at every $k$-point in the Brillouin zone considering the contributions from all possible electronic states. The value of the complex refractive index $\tilde{n}$ of BP was then calculated for the zigzag, out-of-plane and armchair directions. For BP, the complex refractive index is expressed as $\tilde{n_{j}} = \sqrt{\eta_{j} + i\kappa_{j}}$ with $\eta$ and $\kappa$ the real and imaginary parts, respectively where the index $j$ refers to the crystallographic direction.

\section{Raman intensity correction factor}

The intensity correction factor was calculated following a method reported previously \cite{Yoon2009}. The correction factor of the incident $F_{ab}$ and scattered $F_{sc}$ light caused by multiple reflections at a position $x$ measured from the BP surface are given by:
\begin{align}
	F_{ab}=t_{1}e^{-i \beta_{abs} x} \\
	F_{sc}=t'_{1}e^{-i \beta_{sc} x}
\end{align}

\noindent
where $t_{1}= 2n_{0}/(\widetilde{n}_{1}+n_{0})$ and $t'_{1}= 2\widetilde{n}_{1}/(\widetilde{n}_{1}+n_{0})$ are the Fresnel transmittance coefficients of the incident and scattered lights, respectively, with $n_{0}$ and $\widetilde{n}_{1}$ the refractive indices of air and black phosphorus respectively. $\beta_{abs} = 2\pi x \widetilde{n}_{1}/\lambda_{abs}$; $\beta_{sc} = 2\pi x \widetilde{n}_{1}/\lambda_{sc}$. Here, given the low energy of $A_{g}$ phonons with respect to the excitation energy, we take $\lambda_{ab} = \lambda_{sc}$.\\

\noindent
The total correction factor is then expressed as:

\begin{equation}
	F=N \int_0^\infty \vert F_{ab}(x) F_{sc}(x) \vert^{2} dx
\end{equation}

\noindent
where $N$ is a normalization constant.\\

\noindent
The intrinsic Raman intensity $I_{j}$ is then expressed as: $I_{j} = I^{0}_{j} / F_{j}$ with $I^{0}$ the measured Raman intensity and $F$ the correction factor.

\end{document}